\documentclass[12pt,showkeys, rotate, epsfig]{article}
\usepackage{epsfig}
\usepackage{graphicx}
\usepackage{amssymb}
\usepackage{amsthm}
\usepackage{amsmath}
\usepackage{amssymb,amsfonts}
\usepackage{psfig}
\usepackage[english,english]{babel}

\newcommand{\een}{\end{subequations}}
\newcommand{\ben}{\begin{subequations}}
\newcommand{\be}{\begin{eqnarray}}
\newcommand{\ee}{\end{eqnarray}}

\makeatletter
\renewcommand\section{\@startsection{section}{1}{0mm}{7mm}{4mm}{\normalfont\large\bfseries\scshape}}

\begin{document}

\begin{titlepage}

\begin{centering}

\hfill hep-th/0507130\\

\vspace{1 in}
{\bf {BULK HIGGS WITH  4D GAUGE INTERACTIONS} }\\
\vspace{2 cm} {A. Kehagias$^{1}$
 and K. Tamvakis$^{2}$}\\
\vskip 1cm
{$^1 $\it{Physics Department, National Technical University\\
15 780 Zografou, Athens,  GREECE}\\
\vskip 1cm
{$^2$\it {Physics Department, University of Ioannina\\
45110 Ioannina, GREECE}}}\\

\vspace{1.5cm}
{\bf Abstract}\\
\end{centering}

We consider a model with an extra compact dimension  in which the
Higgs is  a bulk field while  all other Standard Model fields are
confined on a brane. We find that four-dimensional gauge
invariance can still be achieved by appropriate modification of
the brane action. This   changes  accordingly the Higgs propagator
so that, the Higgs, in all its interactions with Standard Model
fields,  behaves as an ordinary $4D$ field,
although it has a bulk kinetic term and bulk self-interactions.
In addition, it cannot propagate from the brane to the bulk and, thus, no charge
can escape into the bulk but it remains confined on the brane.
 Moreover, the photon remains massless, while
the dependence of the Higgs vacuum on the extra dimension induces
a mixing between the graviphoton and the ${\cal{Z}}$-boson.
 This results in a modification of the sensitive $\rho$-parameter.

 \vfill

\vspace{2cm}

\begin{flushleft}

July 2005
\end{flushleft}
\hrule width 6.7cm \vskip.1mm{\small \small}
\end{titlepage}

\newpage
\section{Introduction.}
The quest for a unification of gravity with the rest of the
fundamental interactions has led to the idea of extra spatial
dimensions, first introduced in  the Kaluza-Klein (KK) theory.
Presently, the only consistent theoretical framework for the
development of such a unified theory is String Theory or some form
of it~\cite{HW}, which also requires extra spacetime dimensions.
In this  higher-dimensional context, gravity describes the geometry of a $D=4+d$ spacetime possessing
 $d$ extra spatial dimensions. Many aspects of particle physics have been considered in this general framework, giving the opportunity of a new
 and fresh look in old and challenging problems.
In particular,  the introduction of D-branes
  has led in a reformulation and reevaluation of
 the original hierarchy problem, by considering compact  internal
 spaces of {\textit{large}} radius $R$,
 possibly corresponding to a fundamental higher-dimensional Planck mass of $O({\rm TeV})$
 ~\cite{AAD}, or even of infinite radius~\cite{RS},\cite{Keh},\cite{KT}. Many other issues  have
 also been reexamined in this general setup, in which, as  a general rule,
  all degrees of freedom of the Standard Model (SM) are assumed to be confined on a
  $4$-dimensional subspace ({\textit{Brane}}), while
  only gravity propagates in the full space ({\textit{Bulk}}).
  The construction of models in which additional fields besides
  the graviton field can propagate in the bulk has been proven to be quite challenging.
It is common wisdom that, in a world with gauge field on
a brane, charged fields may exist only on the brane as a result of gauge invariance. Exception to this
is neutral particles, as for example in brane models with SM
on the brane and a bulk right-handed
neutrino~\cite{DDG1},\cite{Grossman:1999ra},\cite{AH-MR},\cite{TG}.
  The rule is that only fields which are neutral
  under the Standard Model gauge interactions
  can be introduced as bulk fields with brane-interactions.

  In the present article we
  reexamine the question of fields in the bulk.  Considering the Standard Model Higgs,
  we will show that, Higgs fields propagating  in the bulk
but with  {\textit{localized}} gauge interactions  may exist as well.
  We restrict ourselves in the case of
  one extra compact dimension. The Higgs
   is introduced as a bulk field while the gauge bosons and the rest of the Standard
   Model  are strictly confined on the brane.
   We find that the constraint of
    four-dimensional gauge invariance modifies the Higgs propagator so that, the Higgs, in all its interactions
    with Standard Model
     behaves as an ordinary four-dimensional field and although it propagates within the bulk,
    it cannot propagate from the brane to the bulk.
Despite the fact that no charge can escape into the bulk but it remains confined
on the brane, the Higgs has a bulk kinetic term and bulk self-interaction.
    Moreover, a bulk Higgs field can have a
   vacuum expectation value dependent on the extra dimension. Such a
   {\rm VEV}
    induces an interaction between the {\textit{graviphoton}} and the ${\cal{Z}}$-boson.
    This interaction amounts to a mixing of ${\cal{Z}}$ with the graviphoton,
    which has already aquired a mass due to the presence of the brane\cite{SD}.

In the next section 2, we introduce a model with a  bulk 5D scalar
field with 4D $U(1)$ gauge interactions on a brane
 and we present a gauge invariant action.
 In section 3, we discuss the graviphoton-gauge field mixing by a vacuum
 5D scalar field configuration.
 In section 4, we introduce the SM Higgs in the 5D bulk interacting with the rest of the SM fields,
 which are localized on the 4D brane, and we find
 that, although the photon remains massless, there exist a graviphoton-${\cal{Z}}$
 mixing and an associated modification of the $\rho$-parameter.
 In section 5 we discuss quantum effects due to the 5D Higgs, and finally, in section 6, we summarize our findings.

\section{Bulk Higgs and 4D gauge invariance}

    Let us consider a $5D$ space $M^4\times S^1$ with the extra
    compact dimension $x^5=y$ that takes values in the circle $0\leq y\leq R$. A brane is present at the location $y=0$ of this space.
    The five-dimensional metric is taken to the flat metric
    $$G_{MN}=\left(\begin{array}{cc}
    \eta_{\mu\nu}\,&\,0\\
    \,&\,\\
    0\,&\,1
    \end{array}\right)$$
    Consider now a complex scalar field $\Phi(x^{\mu},y)$ with the $5D$ canonical dimension $3/2$.
    In addition to that, there is a four-dimensional $U(1)$ gauge field $B_{\mu}(x)$ propagating on the brane. The action for the model is
    \be{\cal{S}}={\cal{S}}_0+{\cal{S}}_{Br}+{\cal{S}}_{int} , \label{ss}\ee
    where
    $${\cal{S}}_0=-\int\,d^4x\,\int\,dy\,|\partial_M\Phi|^2$$
    is the free-action for the bulk scalar field,
    $${\cal{S}}_{Br}=-\frac{1}{4}\int\,d^4x\,B_{\mu\nu}(x)B^{\mu\nu}(x)\,,$$
    with $B_{\mu\nu}(x)=\partial_{\mu}B_{\nu}(x)-\partial_{\nu}B_{\mu}(x)$is the action for the $U(1)$ gauge field and
    $${\cal{S}}_{int}=-iag\int\,d^4x\,B^{\mu}(x)\left(\Phi^*(x,0)\partial_{\mu}\Phi(x,0)-
\partial_{\mu}\Phi^*(x,0)\Phi(x,0)-igB_{\mu}(x)|\Phi(x,0)|^2\right)\, ,$$
is their interaction.
In the interaction term $g$ is the dimensionless four-dimensional gauge coupling while the parameter $a$ has canonical
dimension $-1$ and corresponds to a length. The action (\ref{ss}) describes a 5D scalar interacting with a 4D gauge field. The scalar field
propagates in the bulk of the space-time, while its interactions are localized on the brane at $y=0$.  Clearly, ${\cal{S}}_{Br}$ is invariant
under the standard $U(1)$ transformation $\delta B_{\mu}(x)=\partial_{\mu}\omega(x)$. However, the bulk
and the interaction Lagrangian are not
gauge invariant as they stand.

At this point let us define the transformation properties of the scalar field. An obvious guess is that the scalar field transforms
only at the position of the brane, i.e., where it experiences the gauge interactions.
Let us consider then the gauge transformations
\be\delta\Phi(x,y)=iag\omega(x)\delta(y)\Phi(x,y)\,\,,\,\,\delta B_{\mu}(x)=\partial_{\mu}\omega(x)\ee
Note that these transformations are singular on the brane, i.e.
$$\delta\Phi(x,0)=iag\omega(x)\delta(0)\Phi(x,0)\, .$$
However, they can be regularized by taking
$a^{-1}=\delta(0)$. Then, we can write\footnote{$$\hat{\delta}(y)=\lim_{\epsilon\rightarrow 0}\left\{\frac{\delta(y-\epsilon)}{\delta(\epsilon)}\right\}\,,\,\,
\,\,\,\hat{\delta}(0)=\lim_{\epsilon\rightarrow 0}\left\{\frac{\delta(0-\epsilon)}{\delta(\epsilon)}\right\}=1$$}
\be
\delta\Phi(x,y)=ig\omega(x)\hat{\delta}(y)\Phi(x,y)\, , \label{gt}
\ee
so that the gauge transformations on the brane are just
$$\delta\Phi(x,0)=ig\omega(x)\Phi(x,0)\,\,,\,\,\delta B_{\mu}(x)=\partial_{\mu}\omega(x)\, .$$
Note that on the brane we can define a four-dimensional field with
the correct canonical dimension
$$\phi(x)=a^{1/2}\Phi(x,0)=\delta(0)^{-1/2}\Phi(x,0)$$
in terms of which ${\cal{S}}_{int}$ is just
$${\cal{S}}_{int}=-ig\int\,d^4x\,B^{\mu}(x)\left(\phi^*(x)\partial_{\mu}\phi(x)-
\partial_{\mu}\phi^*(x)\phi(x)-igB_{\mu}(x)|\phi(x)|^2\right)\, .$$
Under the gauge  transformation (\ref{gt}),  the action (\ref{ss}) transforms as
$$\delta{\cal{S}}
=iag\int\,d^4x\,\omega(x)
\left({\Phi^{\dagger}}''(x,0)\Phi(x,0)-\Phi''(x,0)\Phi^{\dagger}(x,0)\right)\, ,$$
and thus, it is not gauge invariant. However, gauge invariance can be maintained by adding appropriate terms.
Indeed,  let us consider the term
$${\cal{S}}_1=-a\int\,d^4x\,\left({\Phi^{\dagger}}''(x,0)\Phi(x,0)+\Phi''(x,0)\Phi^{\dagger}(x,0)\right)$$
$$=-a\int\,d^4x\,\int\,dy\,\delta(y)\left({\Phi^{\dagger}}''(x,y)\Phi(x,y)+\Phi''(x,y)\Phi^{\dagger}(x,y)\right)\, .$$
It is easy to verify then that
$$\delta{\cal{S}}_1=-iag\int\,d^4x\,\omega(x)\left({\Phi^{\dagger}}''(x,0)\Phi(x,0)-\Phi''(x,0)
\Phi^{\dagger}(x,0)\right)=-\delta{\cal{S}}\, ,$$
and thus,  the total action
$${\cal{S}}_{tot}={\cal{S}}+{\cal{S}}_1$$ is gauge invariant.
The above discussion may easily be generalised to the non-abelian case.

{\section{Bulk Higgs with brane gauge interactions}

Let us consider now a Higgs field $H$ living in a 5D bulk. All its gauge interactions, as well as some of its self-interactions, are localized at $y=0$. Taking into account
gravity with 5D Planck mass $M_5$, we have the action\footnote{$G^{(i)}$ is the determinant of the induced metric on the
 brane. The parameter $T$ is the {\textit{brane tension}} and has canonical dimensions $[T]=4$.}
$${\cal{S}}=M_5^3\int\,d^5x\,\sqrt{-G}\,{\cal{R}}\,-\int\,d^5x\,\sqrt{-G}\left(\,G^{MN}D_MH^{\dagger}D_NH\,+\,V^{(5)}(H)\right)$$
\be -\int\,d^4x\sqrt{-G^{(i)}}\left(T+\,V^{(4)}(H)\,+\,\frac{1}{4}B_{\mu\nu}B^{\mu\nu}\right)\,,\,\ee
where the gauge field strength is $B_{\mu\nu}(x)=\partial_{\mu}B_{\nu}(x)-\partial_{\nu}B_{\mu}(x)$ and the covariant derivative
\be D_{M}\equiv \left\{\begin{array}{l}
D_{\mu}=\partial_{\mu}-iga\delta(y)B_{\mu}(x)\\
\,\\
D_5=\partial_5
\end{array}\right.\ee
Note the presence of both a bulk self-interaction potential $V^{(5)}(H)$ as well as a potential on the brane $V^{(4)}(H)$.
The canonical dimensions of these terms are different, being $[V^{(5)}]=5$ and $[V^{(4)}]=4$.

If we add to ${\cal{S}}$ the extra local term
\be {\cal{S}}_1=-a\int\,d^4x\,\left(\,{H^{\dagger}}''(x,0)H(x,0)\,+\,H''(x,0)H^{\dagger}(x,0)\,\right)\,,\,\ee
the total action ${\cal{S}}+{\cal{S}}_1$ becomes invariant under the set of gauge transformations
\be\,\delta H(x,y)=iga\omega(x)\delta(y)H(x,y)\,\,,\,\,\,\,\,\delta B_{\mu}(x)=\partial_{\mu}\omega(x)\,.\ee

Let us now consider the general KK-form of the 5D metric
\be
G_{MN}=\left(
\begin{array}{cc}
g_{\mu\nu}  +R^2 A_\mu A_\nu\,\,\,& \,\,\,R^2 A_\mu\\
\,&\,\\
R^2 A_\mu\,\,\,&\,\,\, R^2
\end{array}\right)\, , ~~~~~~G^{MN}=\left(
\begin{array}{cc}
g^{\mu\nu}\,\,\, &\,\,\, -A_\mu\\
\,&\,\\
 -A_\mu\,\,\,& \,\,\,R^{-2}+A^\mu A_\mu
\end{array}\right) \, , \label{kk}
\ee
where   the radion field $G_{55}$ has been freezed to  its
{\rm VEV} $R$, the radius of the extra $S^1$ dimension.
With this form of the 5D metric, $g_{\mu\nu}$ describes the 4D graviton and $A_\mu$ the  4D {\textit{graviphoton}}.
In the absence of the brane, the graviphoton $A_\mu$ is a
massless boson corresponding to the translational invariance along $S^1$. However, in the
presence of the brane, translational invariance is broken and the graviphoton becomes massive. This
has been shown explicitly in~\cite{SD} for a fat brane  formed by a kink soliton. This is also the case for a brane with
delta-function profile.
  To see this, let us
 turn off all fields except
gravity. In this case, the 5D action in the presence of the
brane is
\be
{\cal{S}}=M_5^3\,\int\,d^5x\,\sqrt{-G}\,{\cal{R}}\,\,-T\int\,d^4x\,\sqrt{-G^{(i)}}\, ,  \label{act1}
\ee
 where $G_{\mu\nu}^{(i)}$ is the induced metric at the position of the brane
 \be
 G_{\mu\nu}^{(i)}=g_{\mu\nu}+R^2A_{\mu}A_{\nu}\, .
 \ee
Then, we have for the determinant of the induced metric
\be
 &&\det(-G^{(i)})=\det(-g_{\mu\nu})\det(\delta_{\mu}^{\nu}+R^2A_{\mu}A^{\nu})=
 \det(-g)e^{Tr\ln(1+R^2A\otimes A)}\nonumber \\
  && =
 \det(-g)\,e^{\sum_{n=1}^{\infty}\frac{R^{2n}A^{2n}}{n}}=\det(-g)\,e^{\ln(1+R^2A^2)}=
 \det(-g)(1+R^2A_{\mu}A^{\mu})\, . \label{det}
 \ee
Moreover,  with the KK form (\ref{kk}) of the metric, where $g_{\mu\nu}=g_{\mu\nu}(x)$ and $A_{\mu}=A_{\mu}(x)$,
the 5D Ricci scalar ${\cal{R}}$ turns out to be
\be
{\cal{R}}=\overline{{\cal{R}}}-\frac{R^2}{4}F_{\mu\nu}F^{\mu\nu}\, . \label{rkk}
\ee
 $\overline{{\cal{R}}}$ is the 4D Ricci scalar of the 4D metric $g_{\mu\nu}$ and
$F_{\mu\nu}=\partial_{\mu}A_{\nu}-\partial_{\nu}A_{\mu}$ is the field strength of the graviphoton field.
Substituting, (\ref{det},\ref{rkk}) in (\ref{act1}), we get
 \be
{\cal{S}}&=&2\pi R\,M_5^3\,\int\,d^4x\,\sqrt{-g}\,{\cal{R}}+2\pi R^3\,M_5^3\,\int\,d^4x\,\sqrt{-g}\,\left\{-\frac{1}{4}F_{\mu\nu}F^{\mu\nu}\right\}
\nonumber \\
&&
-T'\,\int\,d^4x\,\sqrt{-g}\,(1+R^2A_{\mu}A^{\mu})^{1/2}
\ee
where $T'\equiv T+\langle V^{(4)}+RV^{(5)}+R|\partial_5 H|^2\rangle $ includes the Higgs contribution to the vacuum energy.
Linearizing the above expression, we arrive at the effective 4D action ($2\pi R M_5^3\equiv M_P^2$)
\be
{\cal{S}}&=&M_P^2\,\int\,d^4x\,\sqrt{-g}\,{\cal{R}}-T'\,\int\,d^4x\,\sqrt{-g}\nonumber \\
&&\,+\,(R M_P)^2\int\,d^4x\,\sqrt{-g}\,\left(-\frac{1}{4}F_{\mu\nu}F^{\mu\nu}-\frac{1}{2}
\frac{T'}{M_P^2}\,A_{\mu}A^{\mu}\right)+\dots \label{act2}
\ee
from where we read off the graviphoton mass
$$M_{A}^2=\frac{T'}{M_P^2}=\frac{1}{2\pi}\left(\frac{T'}{RM_5^3}\right)\, .$$
Thus, we see that the brane breaks the $U(1)$ symmetry of the background and gives a non-zero mass to the graviphoton. This mass
 depends on the
brane tension as in the case where the brane is a  kink of finite width, formed by a scalar field~\cite{SD}.

As we are only interested in the Higgs-gauge sector, we may ignore
gravity and the brane tension. In this case, we may  disregard the first two terms in
(\ref{act2}).
The complete action turns out to be
$$
-\,(RM_P)^2\int\,d^4x\, \sqrt{-g}\, \left(\frac{1}{4}F_{\mu\nu}F^{\mu\nu}+\frac{1}{2}
M_A^2\,A_{\mu}A^{\mu}\right)-\int\,d^5x\,\sqrt{-g}\,\,\left(D_{\mu}H^{\dagger}D^\mu H+V^{(5)}\right)$$
$$+R\int\,d^5x\,\sqrt{-g}\,A^{\mu}\left(D_{\mu}H^{\dagger}\partial_5H+\partial_5H^{\dagger}D_{\mu}H\right)
-\int\,d^5x\,\sqrt{-g}\,(1+R^2A_\mu A^\mu)\partial_5H^{\dagger}\partial_5H$$
\be -\int\,d^4x\,\sqrt{-g}\,V^{(4)}-\frac{1}{4}\int\,d^4x\,\sqrt{-g}\,\left(1+R^2A_\kappa A^\kappa\right)B_{\mu\nu}B^{\mu\nu} \label{act3}
\ee
It is obvious that due to the Higgs field, the graviphoton and the brane gauge field are mixed.
We will discuss this mixing and its consequences
below.
For this we define the 4D Higgs field  on the brane
with the correct canonical dimensions
$$\phi(x)\equiv H(x,0)a^{1/2}=H(x,0)\delta(0)^{-1/2}$$
 and the  canonical graviphoton field
$$A_{\mu}(x)\rightarrow (RM_P)^{-1}A_{\mu}(x)\, .$$
Now, let us note  that there are terms quadratic to the gauge and
graviphoton fields coming from the bulk scalar covariant derivatives , as well as a coupling
 $A_{\mu}(D^{\mu}H)\partial_5H^{\dagger}$. We shall assume a $y$-dependent vacuum $H(x,y)$. By setting
 $$Im\left\{\frac{H'(0,0)}{H(0,0)}\right\} \equiv M\, $$
  the vector mass-terms in (\ref{act3}) turns out to be
\be-\int d^4 x \sqrt{-g} \left(g^2 B_\mu B^\mu |\phi|^2+\frac{1}{2}M_A^2 A_\mu A^\mu+2\frac{g \, M}{M_P} A_\mu B^\mu
|\phi|^2\right)\, . \label{q}
\ee
The vacuum is determined from the Higgs field equation
$$\left[\partial_5^2+\mu_0^2-a\left\{\partial_5^2,\,\delta(y)\right\}+\delta(y)\left(\mu_4^2-\lambda_4|H|^2\right)\,\right]H(0,y)=0$$
where, we have taken
$$V^{(4)}=-\mu_4^2|H|^2+\frac{\lambda_4}{2}|H|^4\,\,,\,\,\,\,V^{(5)}=-\mu_5^2|H|^2+\frac{\lambda_5}{2}|H|^4\approx \mu_0^2|H|^2+\dots$$

The above equation posseses solutions of the form\footnote{An alternative solution, with $\mu_0^2<0$, is
$$H(0,y)=C_1\delta(y)+C_2'\sin(|\mu_0|y)$$}
\be H(0,y)=\,C_1\,\delta(y)\,+\,C_2\,\sinh(\mu_0 y)\ee
By substituting, we obtain
$$\nu_0^2+a\delta''(0)-\delta(0)\mu_4^2+\lambda_4\delta^3(0)|C_1|^2=0$$
Since $[\mu_4^2]=1$ and $[\lambda_4]=-2$, we may introduce the canonical $4D$ parameters
$$\overline{\mu}_4^2=a\mu_4^2\,,\,\,\,\,\overline{\lambda}_4=a^2\lambda_4$$
Introducing the renormalized mass $\mu^2\equiv \mu_0^2+a\delta''(0)$, we get
$$|C_1|^2=\frac{a}{\overline{\lambda}_4}\left(\overline{\mu}_4^2-\mu^2\right)$$
The condition
$$\langle H(0,0)\rangle=a^{-1/2}\frac{v}{\sqrt{2}}$$
gives
\be C_1=a^{-1/2}\frac{v}{\sqrt{2}}\,\,,\,\,\,\,\,\frac{v^2}{2}=\frac{\overline{\mu}_4^2-\mu^2}{\overline{\lambda}_4}\,\ee
The graviphoton mixing parameter $M$ introduced above is given by
$$M=\sqrt{2a}\frac{\mu_0}{v}Im\left\{C_2\right\}$$
Taking for simplicity $C_2$ to be purely imaginary, we may rewrite the vacuum solution as
\be H(0,y)=a^{-1/2}\frac{v}{\sqrt{2}}\left(\,a\delta(y)\,+i\frac{M}{\mu_0}\,\sinh(\mu_0 y)\,\right)\,{\label{Higgsvac}}\ee
The solution can be periodic in $y$ by choosing
a purely imaginary $\mu_0$. This is not a problem as the physical Higgs mass, as we will see, is shifted by $a\delta''(0)$.

Denoting by $M_B^2=g^2v^2$ the gauge boson mass in the absence of the graviphoton, we obtain for the mixed mass-matrix of vector bosons
\be
{{\cal{M}}_V^{(0)}}^2=\left(\begin{array}{cc}
M_B^2\,\,&\,\, M_B^2\, \tan \zeta \\
\,&\,\\
 \, M_B^2\, \tan \zeta \,&\,\,M_A^2
\end{array}\right)
\ee
where we have defined
\be
\tan \zeta=\frac{1}{g}\left(\frac{ M}{M_P}\right)
\ee
The mass eigenstates are the gauge fields
\be
&&A_\mu^{(1)}=\cos\xi \,B_\mu\,-\,\sin\xi \,A_\mu\, , \nonumber \\
&&A_\mu^{(2)}=\sin\xi \,B_\mu\,+\,\cos\xi \,A_\mu
\ee
where
\be
\tan 2\xi=\frac{2\tan \zeta }{\left[\left(\frac{M_A}{M_B}\right)^2-1\right]}
\ee
and the corresponding masses of $A_\mu^{(1,2)}$ are
\be
&&M_{(1)}^2=\frac{1}{2}\left(M_A^2+M_B^2+\sqrt{(M_B^2-M_A^2)^2+4M_B^4\tan^2\zeta}\right)  \\
&&\,\,\nonumber\\
&&
M_{(2)}^2=\frac{1}{2}\left(M_A^2+M_B^2-\sqrt{(M_B^2-M_A^2)^2+4M_B^4\tan^2\zeta}\right)
\ee

\section{The Standard Model with a Bulk Higgs.}

Let us consider now the same $5D$ space time $M^4\times S^1$ with a $4D$ brane embedded in it. All Standard Model fields, except the Higgs $SU(2)_L$
doublet, are localized on the brane. The Higgs field experiences the full $5D$ bulk. Nevertheless, gauge interactions are strictly four-dimensional. The same is true for
Yukawa interactions as well. The model is actually a minimal embedding of the Standard Model in extra dimensions with the Higgs doublet as the only field that
lives in the $5D$ bulk.

Ignoring, for the moment, gravity and the graviphoton, the action for this model
may be written as

\be
{\cal{S}}_{SM}=\int \,d^5x \,\sqrt{-g}\,\left(-|D_MH|^2-V^{(5)}(H)-{\cal{L}}_{SM}\delta(y)\right)\,, \label{act}
\ee

where $H(x,y)$ is an $SU(2)_L$ isodoublet with hypercharge $1/2$
\be H=\left(\begin{array}{c}
H^{(+)}\\
\,\\
H^{(0)}
\end{array}\right)\, , \ee
and
the covariant derivatives are given by
\be
D_M=\left\{\begin{array}{lc}
D_{\mu}\equiv \partial_{\mu}-\frac{i }{2}g'\hat{\delta}(y) B_\mu -\frac{i}{2} g\hat{\delta}(y)
 \vec{W}_\mu\cdot\vec{\tau}\, , ~~~~~\mu=0,...3\\
 \,&\,\\
D_5\equiv \partial_5
\end{array}\right.
\ee
The Higgs potential can be taken to be
\be V^{(5)}=-\mu_5^2|H(x,y)|^2+\lambda_5|H(x,y)|^4\,\approx\,\mu_0^2|H(x,y)|^2\,+\,\dots~ .\ee
Localized Higgs self-interactions $V^{(4)}$ of analogous form are present in ${\cal{L}}_{SM}$ as well, namely
\be V^{(4)}=-\mu_4^2|H(x,0)|^2+ \frac{\lambda_4}{2}|H(x,0)|^4\, .\ee

The theory is invariant under the set of {\textit{$4D$-gauge transformations}} of the $SU(2)_L\times U(1)_Y$ gauge group
$$\delta H(x,y)=\frac{i}{2}\hat{\delta}(y)\,\left(\,g\vec{\omega}(x)\cdot\vec{\tau}\, +\,g'\omega(x) \,\right)H(x,y)\,,\,$$
 \be \delta \vec{W}_{\mu}(x)=\partial_{\mu}\vec{\omega}(x)+g\vec{\omega}(x)
\times \vec{W}_\mu(x)\,\,\,,\,\,\,\,\,\,\,\, \delta B_{\mu}=\partial_\mu \omega(x)\,, \label{gtt}
\ee
provided we add to the action the local term
\be
{\cal{S}}_1=- a\int\,d^4x\left({H^{\dagger}}''(x,0)H(x,0)+H''(x,0)H^{\dagger}(x,0)\right)\,.
\ee
The variation of this term cancels the variation
$$\delta{\cal{S}}_{SM}=-\frac{i}{2}a\int\,d^4x\,\left\{\,{H^{\dagger}}''(x,0)\left(g\vec{\omega}\cdot\vec{\tau}+g'\omega\right)H(x,0)-
H^{\dagger}(x,0)\left(g\vec{\omega}\cdot\vec{\tau}+g'\omega\right)H''(x,0)\,\right\}$$
so that the total action ${\cal{S}}={\cal{S}}_{SM}+{\cal{S}}_1$ is $SU(2)_L\times U(1)_Y$-invariant.

The graviphoton mixing to the neutral gauge fields proceeds as before. The relevant mixing terms are
$$
{\cal{S}}_m=
-(RM_P)^2
\int\,d^4x\, \sqrt{-g}\, \left(\frac{1}{4}F_{\mu\nu}F^{\mu\nu}+\frac{1}{2}
M_A^2\,A_{\mu}A^{\mu}\right)
-\int\,d^5x\,\sqrt{-g}\,\,\left(D_{\mu}H^{\dagger}D^\mu H+V\right)\,$$
$$
-\int\,d^5x\,\sqrt{-g}\,(1+R^2A_\mu A^\mu)\partial_5H^{\dagger}\partial_5H+
R\int\,d^5x\,\sqrt{-g}\,A^{\mu}\left(D_{\mu}H^{\dagger}\partial_5H+\partial_5H^{\dagger}D_{\mu}H\right)$$
\be-\frac{1}{4}\int\,d^4x\,\sqrt{-g}\,\left(1+R^2A_\kappa A^\kappa\right)B_{\mu\nu}B^{\mu\nu}
-\frac{1}{4}\int\,d^4x\,\sqrt{-g}\,\left(1+R^2A_\kappa A^\kappa\right)\vec{W}_{\mu\nu}\cdot\vec{W}^{\mu\nu} \label{a2}
\ee
where $B_{\mu\nu},\vec{W}_{\mu\nu}$ are the $U(1)$ and $SU(2)$  field strengths, respectively and $V$ stands for $V^{(5)}+\delta(y)V^{(4)}$.

Let us now consider the Higgs vacuum solution ({\ref{Higgsvac}})
\be
H(0,y)=a^{-1/2}\left(\,a\delta(y)\,+i\frac{M}{\mu_0}\,\sinh(\mu_0 y)\,\right)\langle \phi\rangle
\ee
and introduce a four-dimensional Higgs field isodoublet $\phi(x)$ as
$$\phi(x)\equiv H(x,0)a^{1/2}=H(x,0)\delta(0)^{-1/2}\, .$$
Then, the emerging mass terms for the graviphoton and the gauge vectors turns out to be
\be
{\cal{S}}_{mass}&=&-\int\,d^4x\,\sqrt{-g}\,\left(\,\frac{1}{2}\frac{T'}{M_P^2}\,A_{\mu}A^{\mu}+
\frac{g^2}{4}\vec{W}_{\mu}\cdot \vec{W}^{\mu}|\phi|^2
+\frac{{g'}^2}{4}B_{\mu} B^{\mu}|\phi|^2+\frac{gg'}{2}B^{\mu}\vec{W}_{\mu}\cdot
(\phi^\dag\vec{\tau}\phi)\,\right. \nonumber
\\&&
+g'\frac{M}{M_P}|\phi|^2 A_{\mu}B^{\mu}+g\frac{M}{M_P}A^\mu \vec{W}_{\mu}\cdot(\phi^\dag \vec{\tau}\phi)
+R^2 M^2|\phi|^2A_{\mu}A^{\mu}\Big{)}
\ee
We have introduced a canonically normalized graviphoton field through the rescaling $A_\mu\to A_\mu/(R M_P)$.
Substituting the {\rm VEV}
$$\langle \phi\rangle=\frac{1}{\sqrt{2}}\left(\begin{array}{c}
0\\
\,\\ v
\end{array}\right)$$
we obtain the mass terms
\be
{\cal{S}}_{mass}=\!-\!\int\,d^4x\,\sqrt{-g}\,\left(\frac{1}{2}M_A^2\,A_{\mu}A^{\mu}+
M_W^2W_{\mu}^+{W^-}^{\mu}+\frac{1}{2}M_{Z}^2Z_\mu Z^\mu
-\frac{v M_W}{\cos\theta_W}\left(\frac{M}{M_P}\right)A_\mu Z^\mu\nonumber\right)
\ee
where
\be
M_A^2=\frac{T'}{M_P^2}+ R^2 M^2 v^2\, ,  ~~~~~~
M_W^2=\frac{g^2 v^2}{4}\, , ~~~~~M_Z^2=\frac{M_W^2}{\cos^2\theta_W}\, .
\ee
As usual, we have introduced $\tan\theta_W=\frac{g'}{g}$ and
\be
W_{\pm}^{\mu}=\frac{1}{\sqrt{2}}\left(W_{1}^{\mu}\pm i W_{2}^{\mu}\right)
\,,\,\,\,\,\,Z_{\mu}=\cos\theta_W A_\mu^3-\sin\theta_WB_\mu
\ee
The photon $\sin\theta_WA_{\mu}^3+\cos\theta_WB_{\mu}$ stays massless while the graviphoton is mixed with the neutral $Z$-boson.
 The neutral vector mass matrix is
 \be
{{\cal{M}}}^2=\left(\begin{array}{cc}
M_Z^2\,\,&\,\,-RMvM_Z\\
\,&\,\\
-RMvM_Z\,\,&\,\,M_A^2
\end{array}\right)=\left(\begin{array}{cc}
M_Z^2\,\,&\,\,-\tan\zeta M_Z^2\\
\,&\,\\
-\tan\zeta M_Z^2\,\,&\,\,M_A^2
\end{array}\right)
\ee
and  we have introduced the mixing angle
\be
\tan\zeta\equiv\frac{v}{M_Z}\left(\frac{M}{M_P}\right)\, .
\ee
The mass eigenstates are
\be
&&Z_{1\mu}=\cos\xi Z_\mu-\sin\xi A_\mu\, , \nonumber \\
&& Z_{2\mu}=\sin\xi Z_\mu+\cos\xi A_\mu
\ee
where
\be
\tan 2\xi= \frac{2 \tan \zeta }{ \displaystyle \frac{M_A^2}{M_Z^2}-1}\, .
\ee
The corresponding masses of $Z_1,Z_2$ are
\be
&&M_{Z_1}^2=\frac{1}{2}\left(M_A^2+M_Z^2-\sqrt{(M_A^2-M_Z^2)^2+4\tan^2\zeta \,
M_Z^4}\,
\right)\, , \nonumber\\
&&
M_{Z_2}^2=\frac{1}{2}\left(M_A^2+M_Z^2+\sqrt{(M_A^2-M_Z^2)^2+4\tan^2\zeta \,
M_Z^4}\,
\right)\, .
\ee
For $M_A>>M_Z$, we get
$$M_{Z_1,Z_2}^2\approx \frac{1}{2}\left\{M_A^2+M_Z^2\mp M_A^2\left(1-\left(\frac{M_Z}{M_A}\right)^2+
2\tan^2\zeta\left(\frac{M_Z}{ M_A}\right)^4\right)\right\}$$
$$\approx\left\{\begin{array}{l}
M_Z^2\left(1-\tan^2\zeta\frac{M_Z^2}{ M_A^2}\right)\,+\, \dots\\
\,\\M_A^2\,+\,\dots

\end{array}\right.$$
Identifying $Z_{1\mu}$ with the neutral gauge boson produced at LEP, we may write
$$\frac{M_W^2}{M_{Z_1}^2}\approx \cos^2\theta_W\left(1+\frac{\tan^2\zeta}{\cos^2\theta_W}\left(\frac{M_W^2}{M_A^2}\right)\right)$$
As a result, the graviphoton-Z boson mixing, gives a departure for the Standard Model value ($\rho=1$) of the parameter
$\rho\equiv M_W^2/\cos^2\theta_W M_{\cal{Z}}^2$
\be
\delta \rho =\frac{\tan^2\zeta}{\cos^2\theta_W}\left(\frac{M_W^2}{M_A^2}\right)=\frac{4}{g^2}\left(\frac{M}{M_P}\right)^2\left(\frac{M_W}{M_A}\right)^2
\ee
which is positive.

Moreover, we may consider the neutral current Lagrangian of $Z_\mu$
\be
{\cal{L}}_{NC}=\frac{g}{\cos\theta_W}\frac{1}{2}\sum_i\bar{\psi}_i\gamma_{\mu}\left(T_{3i}(1-\gamma_5)-2Q_i\sin^2\theta_W\right)\psi_i \, Z^{\mu}
\ee
Changing to mass eigenstates, we get
\be
{\cal{L}}_{NC}=\frac{g}{\cos\theta_W}\frac{1}{2}\sum_i\bar{\psi}_i\gamma_{\mu}\left(T_{3i}(1-\gamma_5)-2Q_i\sin^2\theta_W\right)\psi_i \,
 \left(\,\cos\xi\,Z_1^{\mu}\,+\,\sin\xi\,Z_2^{\mu}\,\right)
\ee
which includes a coupling of the physical graviphoton $Z_2^{\mu}$ to matter proportional to $\sin\xi$. For $M_A>>M_Z$, this coupling is of order
$$\xi\approx (\delta\rho)^{1/2}\,\left(\frac{M_Z}{M_A}\right)\, .$$
Then, constraints on the ratio $M/M_A$ may be obtained by considering the $Z_1$-partial width to fermions $Z_1\to f\bar{f}$~\cite{Altarelli:1989qy},
\cite{Langacker:1991pg},
\cite{Babu:1996vt}. Indeed,
with a shift of the $\rho$
parameter, after taking account higher Higgs representations and $m_t$ effects, of order $\lesssim 10^{-3}$~\cite{Eidelman:2004wy},  we get
\be
\frac{M}{M_A}\lesssim 10^{15}\, \Longrightarrow M_A\gtrsim 10^{-15}\,(g\tan\zeta)\,M_P.
\ee
Thus, for $g\tan\zeta\sim O(1)$, we may have
$$M_A>\,10\,TeV\,\,\,.$$

\section{Quantum effects on the brane.}
Radiative processes of Standard Model particles will also, in general, involve virtual bulk fields that interact with them. In the present model
the Higgs field has been introduced as a bulk field and we would expect that its KK excitations will contribute to loop processes on the brane.
 For example, the gauge couplings
are expected to receive contributions from the Higgs that involve the bulk\cite{DDG}. In the simplified $U(1)$ model of section $1$, the lowest order
Higgs contributions to the vacuum polarization are\footnote{The frequencies $\omega_n$ are $2\pi n/R$.}
$$g^2\frac{a}{R^2}\int\,\frac{d^4q}{(2\pi)^2}\,\tilde{A}_{\mu}(q)\tilde{A}^{\mu}(-q)\left\{\int\,\frac{d^4p}{(2\pi)^2}\,\sum_{n}\,\sum_{n'}\,{\cal{D}}(p;\omega_n,\omega_{n'})\right\}\,+\,$$
$$g^2\frac{a^2}{R^4}\int\,\frac{d^4q}{(2\pi)^2}\,\tilde{A}^{\mu}(q)\tilde{A}^{\nu}(-q)\,\int\,\frac{d^4p}{(2\pi)^2}
\,\sum_{n,n',n'',n'''}
(2q_{\mu}+p_{\mu})(2q_{\nu}+p_{\nu})\,{\cal{D}}(p;\omega_n, \omega_{n'})\,{\cal{D}}(q+p;\omega_{n''},\omega_{n'''})$$
where ${\cal{D}}(p;\omega_n,\omega_{n'})$ is the Fourier transform of the Higgs propagator
\be {\cal{G}}(x-x';\,y,\,y')=\frac{1}{(2\pi)^2}\int\,d^4p\,
e^{ip\cdot (x-x')}\frac{1}{R}\sum_ne^{i\omega_n y}\frac{1}{R}\sum_{n'}e^{i\omega_{n'}y'}{\cal{D}}(p;\,\omega_n,\,\omega_{n'})\ee
Notice that only the propagator $G(x-x';\,0,\,0)$ with end points on the brane appears in these graphs. The above sum can be written as
$$g^2 \int\,\frac{d^4q}{(2\pi)^2}\,\tilde{A}_{\mu}(q)\tilde{A}^{\mu}(-q)\,\int\,\frac{d^4p}{(2\pi)^2}\,\overline{\cal{D}}(p)\,\,\,\,\,\,\,\,\,+\,$$
\be g^2\int\,\frac{d^4q}{(2\pi)^2}\,\tilde{A}^{\mu}(q)\tilde{A}^{\nu}(-q)\,\int\,\frac{d^4p}{(2\pi)^2}\,
(2q_{\mu}+p_{\mu})(2q_{\nu}+p_{\nu})\,{\overline{\cal{D}}}(p)\,{\overline{\cal{D}}}(q+p)\label{prop}\ee
in terms of
\be\overline{\cal{D}}(p)=
\frac{a}{R^2}\int\,\sum_{n,n'}\,{\cal{D}}(p;\omega_n,\omega_{n'})=a\int\,\frac{d^4x}{(2\pi)^2}\,e^{-ip\cdot (x-x')}G(x-x';\,0,\,0)\,\ee

The expression (\ref{prop}) is a standard four-dimensional expression for the vacuum polarization. The existence of the fifth dimension is
encoded in the Higgs propagator $G(x-x';\,y,\,y')$. Note however that this is not the free $5D$ propagator,
satisfying\footnote{The parameter $\mu_0^2$ plays the role of an effective
mass.}
$$\left(-\mu_0^2+\partial_{\mu}\partial^{\mu}+\partial_5^2\,\right){\cal{G}}_0(x-x';\,y,\,y')=\delta^{(4)}(x-x')\delta(y-y')$$
In order to maintain four-dimensional gauge invariance, we have introduced the extra local piece ${\cal{S}}_1$ in the brane action.
This has the effect to modify the Higgs propagator. The modified propagator satisfies the equation\footnote{Note that $a\delta(0)=1$.}
\be\,\left[-\mu_0^2+\partial_{\mu}\partial^{\mu}+\partial_5^2-a\left(\partial_5^2\delta(y)+\delta(y)\partial_5^2\right)\,\right]{\cal{G}}(x-x';\,y,\,y')=
\delta^{(4)}(x-x')\delta(y-y')\,\ee
The solution of this equation, although lengthy\footnote{As a shortcut, we may consider the Fourier transform
$${\tilde{\cal{G}}}_1(p;\,y,\,y')=\frac{1}{(2\pi)^2}\int\,d^4x\,e^{-ip\cdot(x-x')}{\cal{G}}(x-x';\,y,\,y')$$
which satisfies the equation
$$\left\{-\mu_0^2-p^2+\frac{\partial^2}{\partial y^2}-a\left(\delta(y)\frac{\partial^2}{\partial y^2}+\frac{\partial^2}{\partial y^2}\delta(y)\right)\,\right\}
{\tilde{\cal{G}}}_1(p;\,y,\,y')=\frac{\delta(y-y')}{(2\pi)^2}$$
and verify by substitution that its solution is
$${\tilde{\cal{G}}}_1(p;\,y,\,y')=-\frac{1}{(2\pi)^2}\left\{\frac{a}{p^2+\mu^2}\delta(y)\delta(y')+K(y-y')-\frac{K(y)K(y')}{K(0)}\right\}$$
where ( $-R/2\leq z\leq R/2$ )
$$K(z)\equiv \frac{1}{R}\sum_{n=-\infty}^{+\infty}\frac{e^{i\omega_n z}}{\overline{p}^2+\omega_n^2}=
\frac{1}{2\overline{p}}\frac{\cosh(\overline{p}(R/2-|z|))}{\sinh(\overline{p}R/2)}$$
$\overline{p}^2\equiv p^2+\mu_0^2$ and $\mu^2=\mu_0^2+a\delta''(0)$.
 The Green's function ${\tilde{\cal{G}}}_1$ is related to $\overline{D}(p)$ as $\overline{D}(p)=a{\tilde{\cal{G}}}_1(p,\,0,\,0)$.}, proceeds in a straightforward fashion and leads to
\be\,\overline{D}(p)=-\frac{1}{(2\pi)^2}\left(\frac{1}{p^2+\mu^2}\right)\ee
with $\mu^2=\mu_0^2+a\delta''(0)$. This is just a free four-dimensional propagator. It is in sharp contrast to the free propagator which should go as
$$-\frac{a}{(2\pi)^3}\frac{1}{R}\sum_n\frac{1}{p^2+\mu_0^2+\omega_n^2}$$
and contains the contributions of the infinite tower of KK states. The constraint of $4D$-gauge invariance has removed all these contributions
from processes on the brane.

We should also note that for $y,y'\neq 0$, we have
$${\tilde{\cal{G}}}_1(p;\,y,\,y')=-\frac{1}{(2\pi)^2}\left\{K(y-y')-\frac{K(y)K(y')}{K(0)}\right\}$$
and the Higgs can propagate within the bulk. However,
in addition to processes on the brane, a Higgs field cannot propagate from the brane to the bulk.
 This is clear from the propagator ${\tilde{\cal{G}}}_1$ for which
$${\tilde{\cal{G}}}_1(p;\,0,\,y')=-\frac{1}{(2\pi)^2}\frac{\delta(y')}{p^2+\mu^2}\, .$$
This is consistent with charge conservation, as charge cannot escape to the bulk through process like, i.e.,  $Z_1\,W^+\to H^+ H^0$.

\section{Brief Summary.}

The present article was motivated by the (im)possibility of
propagation of  fields (scalars or fermions) in the bulk if gauge
fields are localized on a brane. Indeed, it seems that
 gauge invariance requires the presence of gauge fields (photons) in the bulk as well. Then, from the brane-point of
  view, gauge invariance would be violated and charge would escape to the bulk. However, one may try
 to define gauge invariance strictly as a local symmetry of four
  dimensions and keep gauge fields absolutely trapped on the brane. In such a formulation,
gauge invariance turns out to modify the action of the
  charged bulk fields so that, although they can propagate freely within the bulk, they cannot
propagate from the brane towards the bulk. Apart from their
  bulk self-interactions, in all other interactions occurring on the brane they
  behave as ordinary four-dimensional fields. From the point of view of the brane,
  such a theory would be indistinguishable from a theory in which all charged
  fields are confined on the brane. Nevertheless, in the particular example
  of the Standard Model with a bulk Higgs isodoublet field that we have
  considered, a bulk Higgs field opens the possibility of a vacuum expectation value that depends
   on the fifth dimension. Such a VEV induces new interaction terms of
    gravitational origin that can lead to observable results. The five dimensional metric of
   the compactified $M^4\times S^1$ space of the model includes the
   {\textit{graviphoton}} $A_{\mu}(x)$, a massive vector field that owes its mass to
   the breaking of translational
    invariance of the fifth dimension by the brane. The solution ({\ref{Higgsvac}}) for the
    Higgs vacuum allows for the coupling $A_{\mu}(D^{\mu}H)\partial_5H^{\dagger}$. This coupling
    induces a mixing between the graviphoton and the
    massive neutral vector boson of the Standard Model. This coupling modifies slightly the
    mass eigenvalues and the sensitive mass ratio $M_W/M_{\cal{Z}}$.

    The renormalizability of the model is another important issue.
    Although all Higgs interactions localized on the brane are renormalizable,
its bulk self-interactions are not. Nevertheless, this bulk
self-interaction could very well vanish without effect on the
results and we would have a Higgs freely propagating in the bulk
and a renormalizable $5D$    field theory, letting aside gravity,
of course. In addition to open issues such as the above question
about the precise structure of the bulk, there is a number of
interesting possibilities arising from the central ideas of the
model. The most obvious of them is the possibility of promoting
charged fermions into bulk fields. This  and related questions are
presently under investigation and will be the subject of a future
publication\cite{BKT}.

\vspace{.5cm} \noindent {\textbf{Aknowledgements}}\\
 We would like to thank T. Gherghetta for correspondence.
This work is co - funded by the European Social Fund (75\%) and
National Resources (25\%) - (EPEAEK-B') -PYTHAGORAS

\end{document}